\documentclass[a4paper,12pt]{article}
\usepackage{epsfig}
\usepackage{amssymb}
\usepackage{amsfonts}

\newskip\humongous \humongous=0pt plus 1000pt minus 1000pt

\newif\ifdtup

\catcode`\@=11

\@addtoreset{equation}{section}
\def\theequation{\thesection.\arabic{equation}}

\def\@normalsize{\@setsize\normalsize{15pt}\xiipt\@xiipt
\abovedisplayskip 14pt plus3pt minus3pt%
\belowdisplayskip \abovedisplayskip
\abovedisplayshortskip \z@ plus3pt%
\belowdisplayshortskip 7pt plus3.5pt minus0pt}

\def\small{\@setsize\small{13.6pt}\xipt\@xipt
\abovedisplayskip 13pt plus3pt minus3pt%
\belowdisplayskip \abovedisplayskip
\abovedisplayshortskip \z@ plus3pt%
\belowdisplayshortskip 7pt plus3.5pt minus0pt
\def\@listi{\parsep 4.5pt plus 2pt minus 1pt
     \itemsep \parsep
     \topsep 9pt plus 3pt minus 3pt}}

\relax

\catcode`@=12

\catcode`\@=11

\def\section{\@startsection{section}{1}{\z@}{3.5ex plus 1ex minus
   .2ex}{2.3ex plus .2ex}{\large\bf}}

\def\thesection{\arabic{section}}
\def\thesubsection{\arabic{section}.\arabic{subsection}}

\def\appendix{\setcounter{section}{0}
 \def\thesection{Appendix \Alph{section}}
 \def\thesubsection{\Alph{section}.\arabic{subsection}}
 \def\theequation{\Alph{section}.\arabic{equation}}}

\begin{document}

\newcommand{\beq}{\begin{equation}}
\newcommand{\eeq}{\end{equation}}
\newcommand{\bea}{\begin{eqnarray}}
\newcommand{\eea}{\end{eqnarray}}
\newcommand{\beas}{\begin{eqnarray*}}
\newcommand{\eeas}{\end{eqnarray*}}
\newcommand{\defi}{\stackrel{\rm def}{=}}
\newcommand{\non}{\nonumber}
\newcommand{\bquo}{\begin{quote}}
\newcommand{\enqu}{\end{quote}}
\newcommand{\mat}{\mathbf}
\newcommand{\R}{\ensuremath{\mathbb R}}
\newcommand{\C}{\ensuremath{\mathbb C}}
\newcommand{\Z}{\ensuremath{\mathbb Z}}
\renewcommand{\H}{\ensuremath{\textup{H}}}
\newcommand{\lcm}{\ensuremath{\textup{lcm}}}
\def\de{\partial}
\def\Tr{ \hbox{\rm Tr}}
\def\const{\hbox {\rm const.}}
\def\o{\over}
\def\im{\hbox{\rm Im}}
\def\re{\hbox{\rm Re}}
\def\bra{\langle}\def\ket{\rangle}
\def\Arg{\hbox {\rm Arg}}
\def\Re{\hbox {\rm Re}}
\def\Im{\hbox {\rm Im}}
\def\diag{\hbox{\rm diag}}
\def\longvert{{\rule[-2mm]{0.1mm}{7mm}}\,}
\def\Z{\mathbb Z}
\def\N{{\cal N}}
\def\tq{{\widetilde q}}
\def\W{{\cal W}}
\def\tQ{{\widetilde Q}}
\def\dag{{}^{\dagger}}
\def\p{{}^{\,\prime}}
\def\a{\alpha}
\def\Tr{ \hbox{\rm Tr}}
\def\tM{{\widetilde M}}
\def\tm{{\widetilde m}}
\def\T{{\cal T}}
\def\t{T}
\def\J{{\cal J}}


\begin{titlepage}
\begin{flushright}
ULB-TH-05/15\\
hep-th/0506174\\
\end{flushright}

\bigskip

\begin{center}
{\Large

{\bf  Stable vs Unstable Vortices in SQCD}

 }
\end{center}

\renewcommand{\thefootnote}{\fnsymbol{footnote}}
\bigskip
\begin{center}
{\large   Stefano Bolognesi \footnote{\texttt{s.bolognesi@sns.it}}$^{(1,2)}$   and Jarah Evslin\footnote{\texttt{ jevslin@ulb.ac.be}} $^{(3)}$}
 \vskip 0.20cm
\end{center}

\begin{center}
{\it   \footnotesize
Scuola Normale Superiore - Pisa,
 Piazza dei Cavalieri 7, Pisa, Italy $^{(1)}$ \\
\vskip 0.10cm
Istituto Nazionale di Fisica Nucleare -- Sezione di Pisa, \\
     Via Buonarroti, 2, Ed. C, 56127 Pisa,  Italy $^{(2)}$\\
\vskip 0.50cm
International Solvay Institutes,\\
Physique Th\'eorique et Math\'ematique,\\
Universit\'e Libre
de Bruxelles,\\C.P. 231, B-1050, Bruxelles, Belgium $^{(3)}$}
\end {center}

\setcounter{footnote}{0}

\bigskip
\bigskip

\noindent
\begin{center} {\bf Abstract} \end{center}
We give a topological classification of stable and unconfined massive particles and strings (and some instantons) in worldvolume theories of M5-branes and their dimensional reductions, generalizing Witten's classification of strings in SYM.  In particular $4$d $\N=2$ SQCD softly broken to $\N=1$ contains torsion (Douglas-Shenker) $\Z_N$-strings and nontorsion (Hanany-Tong) $\Z$-strings.  Some of the former are stable when the flavor symmetry is gauged, while those that are not stable confine quarks and in some vacua even dyons into baryons.  The nontorsion strings are stable if and only if all colors are locked to flavors, which is weaker than the BPS condition.  As a byproduct unstable string decay modes and approximate lifetimes are found.  Cascading theories have no vortices stabilized by the topological charges treated here and in particular Gubser-Herzog-Klebanov axionic strings do not carry such a charge.

\vfill

\begin{flushleft}

\end{flushleft}

\end{titlepage}

\bigskip

\hfill{}

\section{Introduction}
During the past year a number of objects of have been discovered
in $\N=1$ supersymmetric gauge theories, such as boojums at the
interface of vortices and domain walls \cite{ST,ASY} and an
axionic string \cite{GHK}.  Besides the new axionic string a
number of other strings are known to exist in this theory, such as
the torsion ($\Z_n$) charged generalizations of Douglas-Shenker
strings \cite{DS,Kneipp} in pure SYM, the nontorsion ($\Z$)
charged and BPS strings of
Refs.~\cite{HT,tong-monopolo,HT2,SY-vortici,SakaiVort} and the
non-BPS strings of
Refs.~\cite{Yung:2000uy,VY,MY,vortici,monovortice,stefano} which
are thought to be unstable and so carry no conserved charges. This
wide variety of known strings leads one to wonder whether even
more strings await discovery.

In this note we will use M5-brane constructions of various $\N=1$
and $\N=2$ supersymmetric gauge theories to find a classification
of the conserved charges carried by the unconfined and stable
massive matter in these theories in terms of the topology of the
configuration.  We will consider matter that arises from the
dimensional reduction of M2-branes, although some instantons come
instead from momentum modes about nontrivial cycles in the
spacetime.  M2-branes can only end on M5-branes, and so,
generalizing Witten's classification of Douglas-Shenker strings in
pure super Yang-Mills \cite{witten-n=1,HSZ}, matter charges
correspond to worldvolumes of M2-branes with boundaries on the
M5-brane.  We will argue that the topological charges are valued
in the relative homology groups $\H_k(M,\Sigma)$ where $M$ is the
internal spacetime manifold, $\Sigma$ is the embedding of the M5
in the internal space, and $k=1,\ 2$\ and $3$ for strings,
particles and certain instantons respectively.  If a string
carries a conserved charge then there is no matter in the theory
on which that string can end, and so the corresponding string
charge will necessarily not be screened.  There may be matter on
which two strings carrying conserved charges can end, but such
matter does not screen the charge of the individual strings,
instead it screens the difference in charges of the two strings.

The charge groups will first be calculated for pure super Yang-Mills with a polynomial superpotential that breaks the gauge symmetry from $U(N)$ to a product of $U(N_i)$'s.  Here we will see that the only topologically stable (not screened) strings are Douglas-Shenker strings which are charged under a single torsion group $\Z_K$ where $K$ is the confinement index of Ref.~\cite{indicediconfinamento}.  In addition there will be various 't Hooft-Polyakov monopoles and dyons, as well as W bosons which will usually be confined.  The fact that these particles are confined by torsion charged strings means that there will be finite combinations of particles that will be unconfined, leading to a rich spectrum of stable electric, magnetic and mixed glueballs with masses proportional to the various Higgs VEVs.

When fundamental matter is added the Douglas-Shenker strings become unstable, they may decay via the nucleation of quark-antiquark pairs and correspondingly the relative homology group that classifies them vanishes.  The strings may still be quite long-lived if a large bare mass is given to the quarks, and they are still physically relevant as they confine the quarks into baryons.  One may try to save some Douglas-Shenker strings by making some of the bare quark masses degenerate, in which case the relative homology group becomes nontrivial, or more precisely becomes dependent upon the way the topology at infinity is treated.  However under an arbitrarily small perturbation this degeneracy is destroyed, and so we claim that the strings may decay.  That is, the physics determines the kind of homology used.  On the other hand one may make the degeneracy stable against small fluctuations by gauging the flavor symmetry, and we will see that in some such theories there are vacua with stable Douglas-Shenker strings.  However in cascading theories and more generally in baryonic vacua it appears as though the corresponding homology group is trivial and so no strings carry the topological charge classified by this group.

In the presence of fundamental matter there are supersymmetric configurations in which the M5-brane may have two connected components, or even more for some theories that in the UV already have several gauge groups.  We will see that the homology group classifying strings contains a number of $\Z$ factors equal to the number of components minus one.  For example, in SQCD with a FI term one finds, as has been demonstrated already in Refs.~\cite{HT,tong-monopolo} in three and four dimensions, stable BPS vortices.  If one introduces a superpotential then these vortices are generically no longer BPS, although supersymmetries may appear in the worldsheet theory that were not present in the bulk theory \cite{ubersizeme}.  However for any superpotential polynomial in the adjoint chiral multiplets the M5-brane will remain disconnected if the FI term is nonzero, and so even when they are non-BPS we claim that these vortices will be absolutely stable.

On the other hand vortices of the type studied in
Refs.~\cite{Yung:2000uy,VY,MY,vortici,monovortice,stefano}
correspond to connected M5-branes and so may decay via
monopole-antimonopole creation.  An FI term is not required for
the stability of these vortices.  Instead we will see that
vortices created from superpotentials alone will be stable
whenever every color is locked to a flavor by a nonvanishing meson
VEV, or equivalently when all monopoles are confined by either 0
or 2 vortices.  We will also argue that the axionic vortices of
Ref.~\cite{GHK} come in two varieties and both are always
unstable.

In this analysis we treat the spacetime and the M5-brane embedding classically.  This supergravity limit may be different from the limit in which the dimensionally-reduced theory of interest is obtained.  Thus calculations of tensions and lifetime should be independently verified in the theories in question when they are not protected by nonrenormalization theorems.  However the topological charges computed in this note, at least in the examples that we know, survive the change of limits despite the fact that many of the stable objects are not BPS.

In section~\ref{caricasezione} we argue that in general the relative homology of the M5-brane embedding is a group of conserved charges and we review an exact sequence which will be used repeatedly throughout the paper to calculate the relative homologies in examples.  In section~\ref{4sezione} we describe a simple example, M-theory compactified on a 2-torus on which three M5-branes are wrapped.  This gives a dimensional reduction of 5-dimensional $U(3)$ pure super Yang-Mills, which consists of $\N=4$ 4-dimensional $U(3)$ super Yang-Mills coupled to a $U(3)$ lower-form gauge theory with a zero-form connection and a one-form field strength.  The various M2-branes connecting the M5-branes correspond to the stable objects known to exist in the gauge theory plus those of the 1-form theory.  Next in section~\ref{2sezione} we extend this analysis to $\N=2$ pure super Yang-Mills, or more precisely flavorless MQCD, by compactifying M-theory on a single circle and considering an M5-brane embedding given by the logarithm of the corresponding Seiberg-Witten curve \cite{witten-n=2}.  A superpotential is added, softly breaking the supersymmetry to $\N=1$, in section~\ref{1sezione} and a number of examples are considered.  Finally in section~\ref{saporisezione} flavored matter is added, and both global and local flavor symmetries are considered.  Gubser-Herzog-Klebanov strings are described in the IIA theory, but the calculation of conserved vortex charges shows that they do not carry any conserved charge of the kind classified here.

\section{Charges from Relative Homology} \label{caricasezione}

Consider M-theory on an 11-dimensional spacetime of the form $\R^{1,d}\times M$ where the gauge theory of interest lives on the spacetime $\R^{1,d}$ and $M$ will be referred to as the internal space.  All of the results in this paper extend equally well to the case in which $\R^{1,d}$ is replaced by an arbitrary ($d$+1)-dimensional manifold.  We will be interested in the low energy theory of an M5-brane that fills the gauge theory $\R^{1,d}$ and sweeps out a $(5-d)$-manifold $\Sigma\subset M$.  In general $\Sigma$ and even $M$ may depend upon the position in the physical spacetime $\R^{1,d}$, for example SQCD and SYM contain domain walls that separate regions with topologically distinct embeddings.  In this note we will consider a region of spacetime in which the topology of the embedding of the M5 in the internal space is constant.

We will now describe a classification of topological charges carried by M2-branes whose boundaries lie entirely on the M5-brane.  M2-brane boundaries can only lie on M5-branes because the supergravity 7-form current $*G_4+C_3\wedge G_4$ is gauge-invariant and in particular globally defined.  This means that it is annihilated by the square of the exterior derivative.  Identifying $dG_4$ and $d*G_4$ with the M5-brane and M2-brane charge densities $\rho_5$ and $\rho_2$ respectively the gauge-invariance of the 7-form implies
\beq
0=d^2(*G_4+C_3\wedge G_4)=d\rho_2+G_4\wedge\rho_5+C_3\wedge d\rho_5\label{termini}
\eeq
where the gauge invariance of $G_4$ implies that $d\rho_5$ vanishes.  The boundary of an M2-brane is characterized by the nonvanishing of $d\rho_2$, but Eq.~(\ref{termini}) implies that when $d\rho_2$ is nonzero $\rho_5$ is also nonzero and so there must be an M5-brane at the M2's boundary.  Of course this argument does not apply at the end of the world where derivatives are not defined, but we will restrict our attention to spacetimes with no boundaries.

If the spacetime is geometrically a product of $\R^{1,d}\times M$, as it will be for the gauge theories of interest far from a domain wall, we may deform the worldvolume of each M2-brane into a number of components each of which is a product of a submanifold of the internal space and a submanifold of the gauge theory spacetime.  For example a diagonal line can be deformed into a straight horizontal and a straight vertical line without changing its topology.  We will see that such composites of products correspond to composite objects in the gauge theory, for example, monopoles with vortices attached.  We will then classify each product manifold separately.

Strings of tension $T$ correspond to M2-branes that extend in two gauge theory spacetime directions while the third is a line segment in $M$ of length $T$ bounded by the M5-brane.  Similarly particles of mass $M$ correspond to M2-branes that extend in one gauge theory spacetime direction while the other two sweep out an area $M$ surface bounded by a collection of curves in the M5.   M2-branes that correspond to instantons are extended in no gauge theory spacetime directions and an internal 3-manifold bounded by a collection of surfaces in the M5.

However there are also composite configurations, in which, for example, the 2-dimensional sheet corresponding to a particle has one boundary not on the M5.  Such a boundary is a 1-dimensional line corresponding to a vortex which, as the M2 is everywhere 3-dimensional, continues into a spacetime direction.  That is, while the M2 can only end on the M5 it may, now that we have distorted it into product submanifolds, have corners in which it turns into a spacetime direction.  In the dimensionally reduced theory this configuration appears to be a particle with a vortex ending on it, in other words this particle is confined and the vortex charge is screened.  In general a particle is confined by a number of vortices equal to the number of components of the boundary of the corresponding sheet that are not on the M5-brane, which in SQCD may be 0, 1 or 2 \cite{noi3}.  We are classifying unconfined objects, and so we are searching for a group of charges that corresponds to surfaces with boundaries only on the M5.

\begin{figure}[ht]
\begin{center}
\leavevmode
\epsfxsize 14   cm
\epsffile{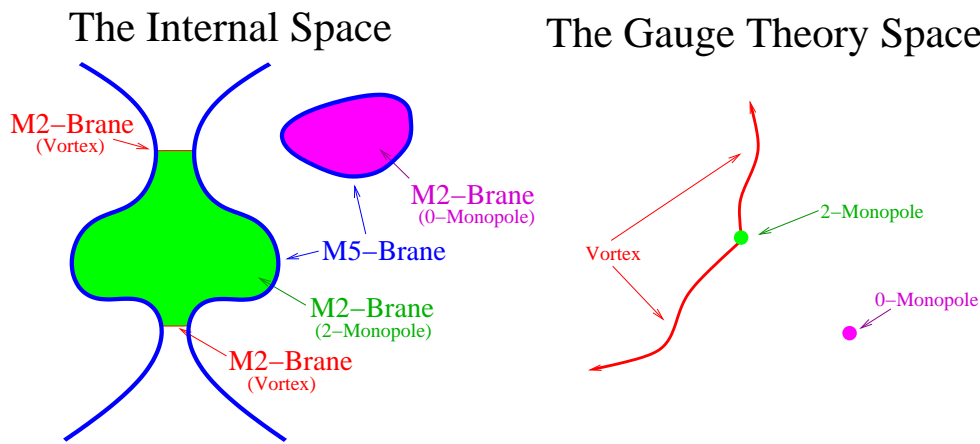}
\end{center}
\caption{ \footnotesize
One M2-brane is bounded by an M5, while another has two boundary components on the M5 but also extends to infinity in two directions.  In the internal space (left) the former is a 2-dimensional sheet surrounded by the M5, while the latter consists of a two-dimensional sheet bounded by the M5 and also two line segments.  In the spacetime of the gauge theory (right) the first M2 is an unconfined monopole while the second is a monopole confined by two vortices. Only the first corresponds to a nontrivial relative homology class.}
\label{confinamento}
\end{figure}

If a given set of vortices confines a particle then this set of vortices is unstable and can decay via nucleation of that particle and its antiparticle.  The lifetime is exponential in the mass of the particle, so such vortices may last for quite awhile, but because their lifetime is finite they will carry no conserved charge.  Similarly instantons may be ``confined'' by particles if the corresponding 3-manifolds have boundaries off of the M5, and the particles may decay via the confined instanton.  Thus the group of conserved charges in dimension $k$ consists of the k-manifolds with boundaries lying entirely along the M5 (the unconfined) quotiented by those that are themselves boundaries (the unstable).  This is the definition of the relative homology group $\textup{H}_k(M,\Sigma)$.

The group $\H_3(M,\Sigma)$, which classifies instantons, is trivial for 4-dimensional SQCD, where $\Sigma$ is the logarithm of the corresponding Seiberg-Witten curve and $M=\R^6\times S^1$.  However if this theory is dimensionally reduced on a spacetime circle then the particles of the 4-dimensional theory may wrap this circle and their dimensional reductions will be instantons that carry a charge classified by $\H_3(M,\Sigma)$.  Such wrapping configurations need not be instantonic, but may also correspond to a particle-antiparticle pair that is created, circumnavigates the circle, and then annihilates leaving only a quantized flux residue behind.  If the particle was confined by a string in the 4-dimensional theory, then the dimensional reduction of the string to 3-dimensions will be an unstable particle that decays by the dimensional reduction of the above process.

The relative homology groups will be calculated using the long exact sequence for relative homology
\begin{equation} \label{seq}
\cdots\stackrel{j^{k+1}_*}{\longrightarrow}\H_{k+1}(M,\Sigma)\stackrel{\partial^{k+1}_*}{\longrightarrow}\H_k(\Sigma)\stackrel{i^k_*}{\longrightarrow}\H_k(M)\stackrel{j^k_*}{\longrightarrow}\H_k(M,\Sigma)\stackrel{\partial^k_*}{\longrightarrow}\cdots
\end{equation}
where $i_*$ is the map induced by the inclusion $i:\Sigma\longrightarrow M$, $j$ is the quotient on chains by the image of the M5-brane, and $\partial_*$ is the map induced by the boundary map.

\section{$\N=4$ U(3) Gauge Theory With Strings} \label{4sezione}

A 4-dimensional $\N=4$ $U(3)$ gauge theory coupled to a lower-form gauge theory with a zero-form $U(3)$ connection and one-form field strength may be engineered from M-theory on $\R^{1,8}\times T^2$ with 3 M5-branes wrapping the $T^2\subset\R^5\times T^2$ and extended along $\R^{1,3}$.  Matter corresponds to M2-branes extending between any two of the M5's and wrapping some subtorus of the $T^2$.  The ordinary gauge theory dyons come from M5's wrapping a circle $S^1\subset T^2$ and extending along a strip whose boundary consists of a line, the particle trajectory, along two M5's.  There are also M2-branes that wrap the entire torus and a line segment between two branes.  These are the electrically charged instantons of the one-form theory.  Finally there are M2-branes that do not wrap the torus at all but fill a 3-dimensional strip which is a line segment between two M5's crossed with a 2-dimensional surface along the M5's.  This 2-dimensional surface is the worldsheet of the 't Hooft-Polyakov string which is magnetically charged under the abelian gauge 1-form field strength.  Note that the particles are not mutually BPS with the strings and instantons, although the strings and instantons may be mutually BPS with each other.

The masses of the particles, tensions of the strings and actions of the instantons are proportional to the corresponding Higgs VEVs, which are the distances between pairs of M5's multiplied by the volumes of the wrapped subtorii. The instanton, particle and string each transform in the adjoint of U(3) and correspondingly each has 6 massive components corresponding to the roots of U(3), two of which are independent corresponding to the two simple roots.  Thus the charge group is $\Z^2$ for the instanton and string, and $\Z^4$ for the particle since there will be a W boson and also a monopole for each simple root.

If two of the M5's are coincident then some of the objects become massless, but under a slight deformation they will be massive.  The classification scheme proposed in this note only applies to configurations that are stable under slight deformations, otherwise, for example, we would have predicted the existence of stable Douglas-Shenker strings in the case in which two M5's are degenerate.  Homology classes that are unstable under small perturbations of the configuration do not correspond to stable objects, instead these objects delocalize out of existence as one approaches transitions in the moduli space.  Thus we will restrict our attention to the case in which no two M5's are coincident.

The relative homology groups are easily computed from the exact sequence (\ref{seq}).  The homology of the internal space $M$ is just that of the torus $T^2$, since the remaining $\R^5$ is contractible.
\begin{equation}
\H_0(M)=\H_2(M)=\Z,\qquad\H_1(M)=\Z^2.
\end{equation}
The M5 wraps the torus, and so the homology of each component of the M5 will also be that of the torus, and that of the whole M5 will then be that of the torus cubed
\begin{equation}
\H_0(\Sigma)=\H_2(\Sigma)=\Z^3,\qquad\H_1(\Sigma)=\Z^6.
\end{equation}
Each M5 component wraps the torus once, and so the $i_*$ maps will be a copy of the identity on each component.  In particular, on $\H_0$ and $\H_2$ it will be a 3 by 1 matrix with all entries equal to 1 while on $\H_1$ it will be a $6$ by $2$ matrix made of three copies of the $2$ by $2$ identity identity matrix.  $j_*$ is the zero map as it quotients chains on the torus by themselves.

This leaves three short exact sequences
\begin{equation}
0\stackrel{j^{k+1}_*}{\longrightarrow}\H_{k+1}(M,\Sigma)\stackrel{\partial^{k+1}_*}{\longrightarrow}\H_k(T^2)^3\stackrel{i^k_*}{\longrightarrow}\H_k(T^2)\stackrel{j^k_*}{\longrightarrow}0
\end{equation}
which yields
\begin{equation}
\H_{k+1}(M,\Sigma)=\H_k(T^2)^2.
\end{equation}
In particular
\begin{equation}
\H_{1}(M,\Sigma)=\H_{3}(M,\Sigma)=\Z^2,\qquad\H_{2}(M,\Sigma)=\Z^4
\end{equation}
matching the expectations from the field theory.

We have only classified topologically stable objects charged under the adjoint of the gauge group, but the bulk fields lead to additional objects that are $U(3)$ singlets.  For example the first compactified circle leads to a $U(1)$ gauge symmetry on the remaining 5-dimensional theory, complete with magnetic and electric charges which are realized by D6 and D0-branes respectively.  Curiously these D0-branes, the $U(1)$ electric charges, are also the instantons of the $U(3)$ gauge theory.  Reducing on the second circle leads to a second $U(1)$ gauge symmetry with a new set of electric and magnetic charges, while the old $U(1)$, similarly to the $U(3)$, is decomposed into an ordinary two-form Maxwell theory plus an abelian one-form theory.  The electric charges of the one-form theory are now the instantons of the $U(3)$ gauge theory.  Each D6-brane that does not wrap the second circle, at least in BPS configurations, leads to a flavor of quark matter in the theory.  The other branes do not lead to objects that exist in the pure gauge theory, but rather to topological charges characteristic of one-form and mixed one-form-two-form theories.  While these objects all carry topological charges, they do not correspond to M2-branes ending on the M5-brane and so are missed by the charge classification scheme of this note.

\section{$\N=2$ Super Yang-Mills} \label{2sezione}

We next turn our attention to $\N=2$ super Yang-Mills with gauge group $U(N)$, whose bare Lagrangian in $\N=1$ superspace takes the form
\beq
\label{n=2}
{\cal L}=\int d^2\theta \,\frac{1}{2e^2}\Tr_{N_c}\,(W^{\alpha}W_{\alpha}) +h.c. + \int d^2\theta d^2\bar{\theta}  \,\frac{2}{e^2}\Tr_{N_c}\,(\Phi\dag e^V\Phi e^{-V}).
\eeq
Here $\Phi$ and $W$ are chiral and vector superfields respectively, both transforming in the adjoint of the gauge group.

Following the construction in Ref.~\cite{witten-n=2} this theory is a sector of the theory obtained by embedding an M5-brane in the spacetime $\R^{1,9}\times S^1$ such that the M5 covers the physical spacetime $\R^{1,3}$ and the embedding in the internal $\R^{6}\times S^1$ is given by the logarithm of the Seiberg-Witten curve \cite{SW1}
\beq \label{sw}
y^2=P_N^2(v)-\Lambda^{2N}
\eeq
where $P_N$ is a polynomial of order $N$
\beq
P_N(v)=\frac{1}{2}\prod_{i=1}^{N}(v-\phi_k)
\eeq
constructed from the eigenvalues $\phi_k$ of the adjoint scalar.

The internal space $M$ is $\R^6\times S^1$, which we parametrize with three complex coordinates $v,\ w$\ and $s$ and one real coordinate $x^7$.  $v$, $w$\ and $x^7$ parametrize the $\R^5\subset\R^6\times S^1$ while the complex coordinate $s=x^6+ix^{10}$ parametrizes the remaining $\R\times S^1$.  $x^10$ is a periodic coordinate parametrizing the M-theory circle.  For future use we define the complex coordinate $t=exp(s)$ which is valued in $\C^*$.  These coordinates are related to the Seiberg-Witten curve (\ref{sw}) via the change of coordinates $t=y+P_N(v)$.  In all the M5-brane embedding is the Riemann surface in $M$ given by the equations
\beq
t^2-2P_N(v)t+\Lambda^{2N}=w=x^7=0. \label{n2embed}
\eeq
Dimensionally reducing to type IIA this configuration becomes two parallel NS5-branes with $N$ D4-branes stretched between them.  The effective worldvolume theory of the D4's contains the gauge theory of interest.

The $\N=2$ theory enjoys a $(2N-2)$-dimensional moduli space of vacua in which the dyon masses are smooth functions of the coordinates.  In particular configurations with massless dyons are codimension two in the moduli space and so, as the moduli space is connected, will not satisfy our stability criteria.  Thus it will suffice to consider vacua in which all dyons are massive.  In such vacua the M5-brane consists of $N$ separated tubes which each wrap the M-theory circle once.  All of the tubes connect at large and small $x^6$ to form a genus $N-1$ Riemann surface $\Sigma_{N-1}$ with two punctures at $x^6=\pm\infty$.  The integral homology groups of this Riemann surface are
\begin{equation}
\H_0(\Sigma_{N-1})=\Z,\qquad \H_1(\Sigma_{N-1})=\Z^{2N-1},\qquad \H_2(\Sigma_{N-1})=0
\end{equation}
where the $2N-1$ generators of $\H_1$ are the $N$ $A$-cycles that wrap the tubes and the $N-1$ dual $B$-cycles that run down the $i$th tube and back up the $i+1$st.

The relative homology groups may be calculated using the long exact sequence
\bea \label{n2seq}
0&=&\H_2(\R^6\times S^1)\stackrel{j^2_*}{\longrightarrow}
\H_2(\R^6\times S^1,\Sigma_{N-1})\stackrel{\partial^2_*}{\longrightarrow}
\H_1(\Sigma_{N-1})=\Z^{2N-1}\nonumber\\
&\stackrel{i^1_*}{\longrightarrow}&
\H_1(\R^6\times S^1)=\Z\stackrel{j^1_*}{\longrightarrow}
\H_1(\R^6\times S^1,\Sigma_{N-1})\stackrel{\partial^1_*}{\longrightarrow}
\H_0(\Sigma_{N-1})=\Z\nonumber\\
&\stackrel{i^0_*}{\longrightarrow}&
\H_0(\R^6\times S^1)=\Z\stackrel{j^0_*}{\longrightarrow}
\H_0(\R^6\times S^1,\Sigma_{N-1})=0\ .\nonumber
\eea
The inclusion $i^0_*:\Z\longrightarrow\Z$ is degree one, as the embedding maps a point to a single point, and so the kernel of $i^0_*$ is trivial.  The kernel of $i^0_*$ is the image of $\partial^1_*$ and so $\partial^1_*$ is the zero map and $j^1_*$ is onto
\beq
i^0_*=1,\qquad \partial^1_*=0,\qquad \H_1(\R^6\times S^1,\Sigma_{N-1})=\textup{Image}(j^1_*)\ .
\eeq
Of the $2N-1$ cycles that generate $\H_1(\Sigma_{N-1})$, the $N$ $A$-cycles each wrap the M-theory circle, which generates $\H_1(\R^6\times S^1)$, while a basis of $N-1$ $B$-cycles is chosen such that no $B$-cycle wraps the M-theory circle.  Therefore $i^1_*$ is a $(2N-1)$-dimensional column vector consisting of $N$ $1$'s and $N-1$ $0$'s.  In particular, every element of $\H_1(\R^6\times S^1)$ is in the image of $i^1_*$, reflecting the fact that there is some cycle in the Riemann surface that wraps the M-theory circle any given number of times.  Thus $i^1_*$ is onto, so $j_*^1$ must be the zero map.  However we have seen the $j_*^1$ is onto, thus the group of vortex charges vanishes
\beq
i^1_*(a_i,b_i)=\sum_{i=1}^Na_i,\qquad j^1_*=0,\qquad \H_1(\R^6\times S^1,\Sigma_{N-1})=\textup{Image}(j^1_*)=0
\eeq
and there are no topologically stable vortices in the pure $\N=2$ theory.

The map $j_*^2$ has a trivial domain, and so it has a trivial image.  Thus the map $\partial_*^2:\H_2(\R^2\times S^1,\Sigma_{N-1})\longrightarrow\Z^{2N-1}$ is into.  The image of $\partial_*^2$ is the kernel of $i^1_*$, which consists of all $N-1$ $B$-cycles plus the $(N-1)$-dimensional subspace of $A$-cycles that wrap the M-theory circle zero times.  The fact that $\partial_*^2$ is into implies that this image is also its domain, the group of particle charges $\H_2(\R^2\times S^1,\Sigma_{N-1})$.  Thus there are $2N-2$ independent M2-brane worldvolumes yielding particles, of which $N-1$ are bounded by the $B$-cycles while $N-1$ are bounded by pairs of $A$-cycles with opposite orientations.  These are just a basis of the 't Hooft-Polyakov monopoles and the W bosons.  The entire group of conserved charges,
\bea
\H_2(\R^2\times S^1,\Sigma_{N-1})=\Z^{2N-2}
\eea
consists of all combinations of W bosons and 't Hooft-Polyakov monopoles.  Every such combination has a BPS bound, given by the areas of complex surfaces (when they exist) bounded by the corresponding cycles.  Physically most of these BPS bounds are never saturated, and the various monopoles and W bosons often repel.  However the topological classification is insensitive to this repulsion, it includes non-BPS topologically stable configurations.

Note that there is an analogous configuration to the vortex of the $\N=4$ 1-form gauge theory, an M2 which extends between two tubes but whose boundary consists of two points on the Riemann surface, rather than a cycle. However unlike the $\N=4$ case this configuration does not yield a nontrivial homology class, as it may be pushed in the $\pm x^6$ direction until it is lies on the M5-brane where it may dissolve.  We will argue that this unstable vortex, rotated by 90 degrees, is an $\N=2$ version of the unstable axionic vortex of Ref.~\cite{GHK}.  There is no finite-volume configuration analogous to the $\N=4$ 1-form instanton.  These two configurations, which correspond to the physics of the 1-form theory and not the 2-form gauge theory of interest, reappear in some vacua when we compactify the $x^6$ direction to introduce bifundamental matter. The gauge theory instantons arise from momentum modes about the M-theory circle, and not M2-branes, and so are missed by this classification.

If one adds $N_f\leq 2N-2$ semi-infinite tubes that wrap the M-theory circle, corresponding to the Seiberg-Witten curve
\beq \label{neuno}
y^2=\frac{1}{4}\prod_{k=1}^{N}(v-\phi_k)^2+\Lambda^{2n_c-n_f}\prod_{j=1}^{N_f}(v+m_j)
\eeq
the gauge theory will include $N_f$ flavors of fundamental matter hypermultiplets with bare masses $m_j$.  Correspondingly $\H_1(\Sigma)$ will gain $N_f$ new generators.  $i^1_*$ is already onto, and so these extra generators imply an extra $\Z^{N_f}$ factor in the group of stable particles $\H_2(M,\Sigma)$ corresponding to the quarks.  Linear combinations of all of the particles are now generically flavored dyons.

\section{$\N=1$ Super Yang-Mills} \label{1sezione}

\subsection{The $\N=1$ Curve}

We will now softly break the $\N=2$ supersymmetry to $\N=1$ by introducing a superpotential $\W(\Phi)$ that is degree $m+1$ polynomial in the adjoint chiral multiplet $\Phi$
\bea
\label{sym}
{\cal L}&=&\int d^2\theta \,\frac{1}{2e^2}\Tr_{N_c}\,(W^{\alpha}W_{\alpha}) +h.c. \\ &&+\frac{2}{e^2}\Tr_{N_c}\,(\Phi\dag e^V\Phi e^{-V})+ \int d^2\theta \sqrt{2} \Tr_{N_c}\,\W(\Phi)  +h.c.\ .\nonumber \eea
This corresponds to replacing replacing the embedding condition $w=0$ of Eq.~(\ref{n2embed}) with \cite{deboerdeharo}
\beq
w^2-2\W\p(v)w-\tilde{f}_{m-1}(v)=0 \label{wcond}
\eeq
for some degree $m-1$ polynomial $\tilde{f}_{m-1}$ that captures quantum corrections to the superpotential.

Eq.~(\ref{wcond}) is quadratic and so we have not only deformed but also doubled our original Riemann surface, that is each point $v$ of the $\N=2$ curve has split into two different points representing the two roots of $w(v)$ in (\ref{wcond}).  However in Ref.~\cite{deboerdeharo} the authors have argued that $\N=1$ supersymmetry requires that all odd-degree roots of $w^2$ are also odd-degree roots of $y^2$, and vice versa.  Therefore any loop will encircle an even number of roots, and so change sheets an even number of times.  Thus no path connects the two sheets, and we may throw one sheet away, leaving the Riemann surface undoubled.  If we do not throw this sheet away we will not obtain a deformation of the $\N=2$ theory, but rather two coupled copies of such a deformation.

Reducing to IIA one obtains two NS5-branes, again at different $x^6$ positions.  The NS5 on the left is at $w=0$, while that on the right, which is often named NS5$\p$, is at $w=\W\p(v)$.  The roots of $w(v)$ are thus the critical points of the superpotential, while those of $P_N(v)$ still correspond to the expectation values of the adjoint scalars.  $\N=1$ supersymmetry implies that the roots of $P_N(v)$ are all within a distance of order $\Lambda$ of the roots of $w(v)$.  In field theory this corresponds to the fact that the eigenvalues of the adjoint scalars are at extrema of the superpotential up to quantum corrections of order $\Lambda$.

From the point of view of the brane cartoon this condition reflects the fact that supersymmetry requires the D4's to proceed along the $x^6$ direction without bending in the $w$-plane, and so they may only connect the NS5-branes when they both occupy the same $w$ coordinate.  The NS5 on the left is always at $w=0$, so D4's may only be placed at points $v$ such that the NS5$\p$ is also at $0=w=\W\p(v)$.  The distribution of eigenvalues among the critical points of the superpotential leads to a classical breaking of the gauge symmetry
\beq
U(N)\longrightarrow\prod_{i=1}^{k} U(N_i)
\eeq
where $i$ runs over the $k$ critical points, which are taken to be separated by a distance much greater than $\Lambda$ to avoid exotic vacua such as those of Ref.~\cite{ES}.  Only a finite set of points $\{\phi_k\}$ on the Coulomb branch satisfy the $\N=1$ supersymmetry requirement, corresponding to the existence of $N_i-1$ independent massless dyons in the $U(N_i)$ subsector.  In particular this means that the above stability condition no longer excludes vacua with massless states, as it did in $\N=2$.  In field theory terms this stability is caused by a nontrivial gluino condensate that obstructs the deformations away from these points.  It is the topology of the condensate field that stabilizes the vortex.

The $N_i$ D4-branes at each critical point of the superpotential lift to a single tube of M5-brane which wraps the M-theory circle $N_i$ times.  This tube cannot split into smaller tubes, as the distance between these tubes would give a mass to all of the dyons of their respective gauge groups which would be incompatible with the known gluino condensate.  Thus the $\N=1$ super Yang-Mills M5-brane is a genus $k-1$ Riemann surface $\Sigma_{k-1}$.  However the soliton spectrum is not identical to the $N=k-1$ case of the $\N=2$ curve because the embedding of $\Sigma_{k-1}$ into the internal space $M$ is topologically inequivalent as a result of the fact that the $A$-cycles now wrap the M-theory circle $N_i$ times instead of just once.  In addition the embedding of the $B$-cycles in the spacetime now will sometimes wrap the M-theory circle, in which case the cycle will not bound a disk in $M$ and the corresponding monopole will be confined.  This is in contrast with the $\N=2$ case, where it was always possible to concatenate a $B$-cycle that encircles the M-theory circle $k$ times with $-k$ $A$-cycles to construct a new $B$-cycle with winding number zero.  To understand these phenomena we will consider some special cases.  The case $N_i=1$ is identical to that of $\N=2$.

\subsection{$U(N)$ Super Yang-Mills}

This case has been analyzed by Witten in Ref.~\cite{witten-n=1}.  If, up to corrections of order $\Lambda$, all of the VEVs of the adjoint scalar are at the same critical point of the superpotential, for example if the superpotential is quadratic and so only has one critical point, then the gauge symmetry is classically unbroken, although quantum mechanically it will be dynamically broken.

The Riemann surface is genus zero, with no $B$-cycles and a single $A$-cycle which wraps the M-theory circle $N$ times.  The Witten index of this theory is $N$ and correspondingly there are $N$ distinct Riemann surfaces.  For example in the limit in which the adjoint chiral multiplet mass, which is the quadratic term in the superpotential, goes to infinity one finds
\beq
v^n=t,\qquad w=\zeta v^{-1}
\eeq
where $\zeta$ is an $n$th root of unity that labels the vacuum.  The choice of root of unity corresponds to a choice of how to identify the $N$ $x^{10}=0$ paths on each side of the tube, or more precisely to cyclic permutations of the identifications of the points $x^{10}=0$ at $x^6=+\infty$ and $x^6=-\infty$.  While the choice of identifications will affect the set of topological charges in the later examples, it will not have any effect in the $k=1$ case considered in this subsection.

This Riemann surface is homeomorphic to that of the $U(1)$ case of the $\N=2$ theory, however the embedding is inequivalent as the A-cycle now wraps the M-theory circle $N$ times.  Thus the exact sequence is identical to  Eq.~(\ref{n2seq}) but now $i^1_*$ is multiplication by the number $N$
\beq
i^1_*:\H_1(\Sigma_0)=\Z\longrightarrow\H_1(\R^6\times S^1)=\Z:j\mapsto Nj.
\eeq
In particular $i^1_*$ is no longer onto, the image consists only of integers that are multiples of $N$.

\begin{figure}[ht]
\begin{center}
\leavevmode
\epsfxsize 14   cm
\epsffile{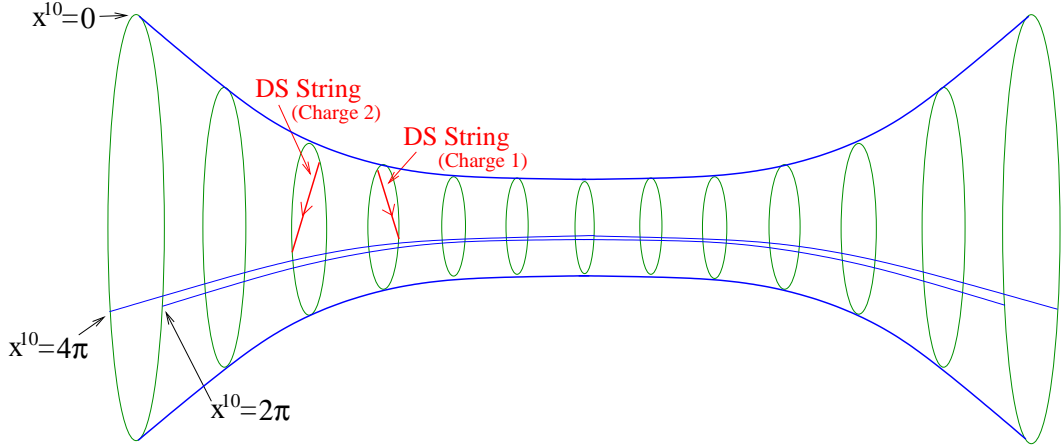}
\end{center}
\caption{ \footnotesize
Pure $U(3)$ super Yang-Mills is a sector of the worldvolume theory of this hourglass-shaped M5-brane.  The M5 has only one $A$-cycle, which winds around the M theory circle 3 times.  Douglas-Shenker strings are M2-branes at constant values of $x^{10}$ that connect distinct points of the $A$-cycle.}
\label{ds}
\end{figure}

Physically this means that an M2-brane that winds around the M-theory circle $N$ times can unwind by touching the M5, opening and letting the two ends slide around the $A$-cycle one time relative to each other.  However a loop that winds less than $N$ times, while it may still open and slide around the $A$-cycle, only changes its winding number by $N$ and so can never get its winding number to zero and disappear.  Thus these winding M2-branes carry stable charges classified by $\Z_N$.

To see this from the exact sequence, recall that $\partial^1_*$ is the zero map.  This implies that the group of conserved vortex charges is
\beq
\H_1(\R^6\times S^1,\Sigma_0)=\textup{Image}(j^1_*)=\frac{\H_1(\R^6\times S^1)}{\textup{Ker}(j^1_*)}=\frac{\H_1(\R^6\times S^1)}{\textup{Image}(i^1_*)}=\frac{\Z}{N\Z}=\Z_N.
\eeq
These are the charges of the Douglas-Shenker strings.  The relative homology groups correspond to lines at constant $x^{10}$ connecting points on the $A$-cycle that are at equal values of the M-theory circle coordinate $x^{10}$.

The rest of the sequence is unchanged and so, as in the $\N=2$
theory, there are no topologically stable, unconfined particles.
For example one may follow a cylinder that was a W boson in the
$\N=2$ $U(N)$ case as the $N$ $A$-cycles merge and one finds that
the two $A$-cycles wrapped by the ends of this cylinder no longer
close \cite{HSZ}.  Instead the circles have been broken, and the
two ends are still at the same $x^{10}$ coordinate but now are
separated by $1/N$ of the $A$-cycle.  While $*G_4+C_3\wedge G_4$
gauge-invariance does not allow an M2-brane to break without all
boundaries laying on the M5, there is a composite
current-conserving configuration that contains this broken
cylinder.  One may attach each broken end of the cylinder to a
strip that continues into one of the gauge theory spacetime
directions.  That is, one may attach the W boson to two
Douglas-Shenker strings.  Thus W bosons have disappeared from the
spectrum because they are confined.  The two strings have opposite
orientations, and so the total string charge emitted from the
confined W boson is zero.  Thus in this example Douglas-Shenker
strings will not be able to decay via the nucleation of pairs of W
bosons, and so their corresponding charges are not screened.  In
fact if a W-boson is inserted in a string worldsheet the $\Z_N$
vortex charge is the same on both sides of the W.  Note that the
relative charges of the two strings that end on a given W-boson
depends on the direction from which the strings approach, as this
determines the relative orientations of the M2-branes in the gauge
theory directions.

\subsection{$U(N+1)\longrightarrow U(N)\times U(1)$ Super Yang-Mills}
This configuration is identical to that of the previous subsection except that now there is an extra tube of M5 that wraps the M-theory circle once.  The $v$ position (adjoint scalar VEV) of this new tube is a different critical point of the superpotential than the point used by the original tube.  Now there will be a second $A$-cycle, which goes around the new tube once, on which $i^1_*$ acts by multiplication by one.

There are now massive W bosons corresponding to cylindrical M2-branes whose two boundary circles, one of which is on each $A$-cycle, each wrap the M-theory circle once.  As in the previous example, the boundary on the $U(N)$ $A$-cycle fails to close, and to make it close a vortex must be inserted.  Thus each W boson is confined by a single $U(N)$ Douglas-Shenker string and in turn each Douglas-Shenker strings may decay via the nucleation of a W-anti-W pair.  The lifetime of such strings is then exponential in the W mass, which is proportional to the distance between the two critical points of the superpotential.  Notice that $N$ W bosons may come together to form an unconfined glueball, as the $N$ vortices confining them may annihilate each other.  This glueball corresponds to a cylindrical M2 whose boundary circles wrap the $U(N)$ $A$-cycle once and the $U(1)$ $A$-cycle $N$ times.  As both winding numbers are integral, both boundaries close.

We may construct the $B$-cycle by connecting the two paths that connect the tubes to paths that run down the two tubes at constant values of $x^{10}$.  A choice of $B$-cycle exists such that $x^{10}$ is constant along the whole cycle and so the $B$-cycle is a boundary in the internal space.  We will refer to the bounded disk as the monopole or $(1,0)$-dyon.

Summarizing, the set of conserved particle charges is identical to the case of a $U(2)$ $\N=2$ theory, although the minimum charge of an unconfined W boson is higher in the $\N=1$ case.  The set of particle charges is generated by the charge 1 monopole and a charge $N$ bound state of W bosons.  No decay modes for particles have been found, and so the particle charge group is expected to be $\Z^2$.  Similarly the vortices can decay via W boson pair production, and so the vortex charge group is expected to be the trivial group $0$.

\begin{figure}[ht]
\begin{center}
\leavevmode
\epsfxsize 10   cm
\epsffile{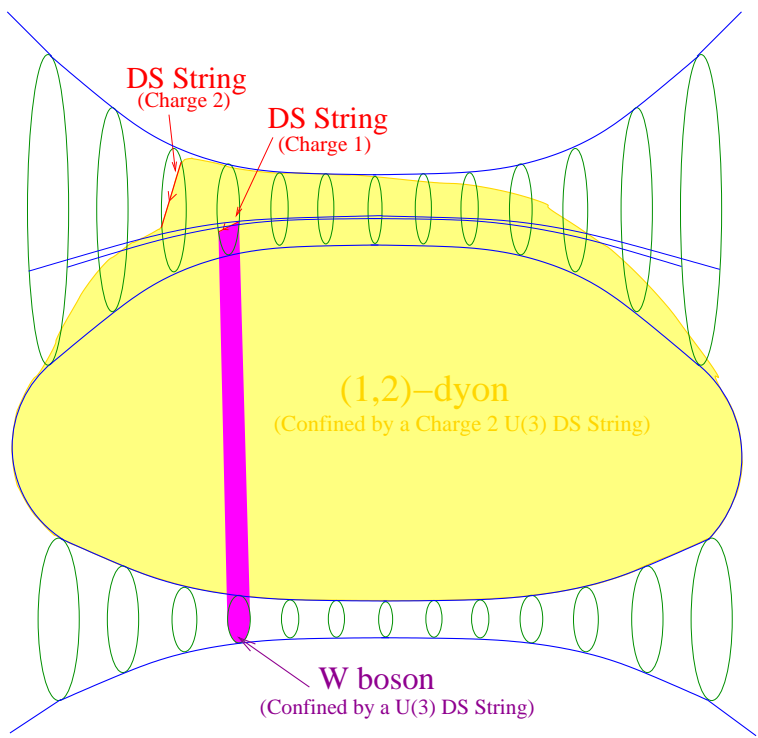}
\end{center}
\caption{ \footnotesize
Pure $U(3)\times U(1)$ super Yang-Mills is a sector of the worldvolume theory of this M5-brane, which is topologically a torus with two punctures.  The M5 has two $A$-cycles, a $U(3)$ cycle that winds around the M theory circle 3 times and a $U(1)$ cycle that wraps once.  There is also a $B$-cycle that does not wrap the M theory circle.  W bosons are confined by the $\Z_3$-charged strings of the $U(3)$. In the $r=0$ vacuum, drawn here, the $(1,2)$-dyon is confined by a charge 2 string and the monopole is unconfined.  In the $r=2$ vacuum the monopole would be confined, and would correspond to the above yellow region.}
\label{ds2}
\end{figure}

This may be compared against the relative homology groups obtained using the long exact sequence.  Everything is as before, except now the inclusion map is
\beq
i^1_*:\H_1(\Sigma_1)=\Z^3\longrightarrow\H_1(\R^6\times S^1)=\Z:(a_1,a_2,b)\mapsto Na_1+a_2.
\eeq
That is, the inclusion map is matrix multiplication by the column vector $(N,1,0)$.  Here $a_1,\ a_2$ and $b$ are the $U(N)$ $A$-cycle, the $U(1)$ $A$-cycle and the $B$-cycle respectively.  As in the $\N=2$ case $i^1_*$ is onto and so $j^1_*$ is the zero map, which again implies that there are no nontrivial topological charges for vortices
\beq
\H_1(\R^6\times S^1,\Sigma_1)=0.
\eeq
The kernel of $i^1_*$ consists of all triplets $(a_1,a_2,b)\in\Z^3=\H_1(\Sigma_1)$ such that $Na_1+a_2=0$.  This is just $\Z^2$, as each choice of $a_1$ and $b$ yields precisely one possible value of $a_2$, namely $a_2=-Na_1$.   $j^2_*$ is the zero map and so $\partial^2_*$ is an isomorphism between this kernel and the particle charge group $\H_2(\R^6\times S^1,\Sigma_1)$.   Therefore the group of conserved particle charges is
\beq
\H_2(\R^6\times S^1,\Sigma_1)=\textup{Image}(\partial^2_*)=\textup{Ker}(i^1_*)=\{a_1,a_2,b|Na_1+a_2=0\}=\Z^2
\eeq
in line with the above field theory expectations.

The $U(N)$ sector has a Witten index of $N$, and so there are $N$ different equivalent vacua, which we will parametrize by a number $r$.  In the $r$th vacuum the $B$-cycle will wrap the M-theory circle $r$ times, so the monopole will be confined by a charge $r$ string.  The $(1,-r)$ dyon, on the other hand, will not be confined, nor will a bound state of $\lcm(r,N)$ monopoles where $\lcm$ is the least common multiple.  The charge group of the strings is independent of the naming convention of the dyons, and so it is the same for all vacua.  In the general case the confinement pattern will depend nontrivially on the choice of $r$. In Ref.~\cite{indicediconfinamento} this $r$-dependent vortex decay via monopole pair creation was refered to as magnetic screening.

\subsection{$U(6)\longrightarrow U(4)\times U(2)$ Super Yang-Mills}
\begin{figure}[ht]
\begin{center}
\leavevmode \epsfxsize 13 cm \epsffile{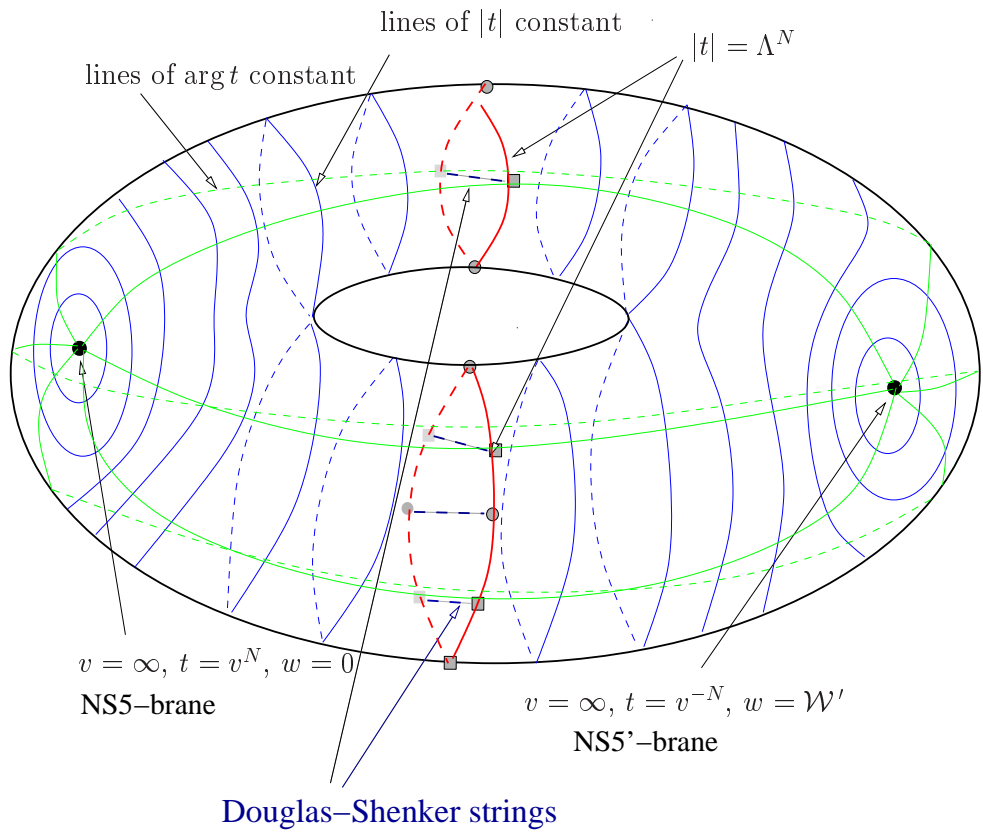}
\end{center}
\caption{ \footnotesize MQCD curve for $U(6) \to U(4) \times
U(2)$.} \label{U(6)}
\end{figure}

Next we will consider a case in which the gauge symmetry is
classically broken from $U(6)$ to $U(4)\times U(2)$ (see Figure
\ref{U(6)}), where the two components reside at two extrema of,
for example, a cubic superpotential.
  As in all $\N=1$ pure
Yang-Mills theories there will be a further dynamical symmetry
breaking to an abelian group, and it is this symmetry breaking
which leads to the existence of discrete vacua.  In this case the
vacua are labeled by two numbers $r_1$ and $r_2$, the first of
which runs from 1 to 4 and the second from 1 to 2.  In addition
there are two kinds of Douglas-Shenker strings, those of the
$U(4)$ and those of the $U(2)$, which we will call 4-strings and
2-strings respectively.

At energies much lower than the separation between the two critical points of the superpotential, this theory reduces to two decoupled pure super Yang-Mills theories with gauge groups $U(4)$ and $U(2)$.  Thus bound states of 4 4-strings or 2 2-strings decay rapidly.  There are W bosons whose masses are proportional to the separation between the critical points.  These correspond to cylinders whose two bounding circles each travel a distance $2\pi$ in $x^{10}$ along one of the two $A$-cycles.  The two $A$-cycles now wrap the M-theory circle $2$ and $4$ times respectively, thus neither of the circles that bounds the cylinder closes.  To make a gauge-invariant M2 configuration we may attach each end of the cylinder to one of the Douglas-Shenker strings of the corresponding $A$-cycle.  Thus the W bosons that are charged under the bifundamental of the $U(4)\times U(2)$ are confined by a 2-string and a 4-string.  This means that the nucleation of a pair of such W bosons may cause a 2-string to be turned into a 4-string, or vice versa.  As a result the charge $(1,1)$ in the maximal vortex charge group $\Z_4\times\Z_2$ is not conserved.  The largest possible conserved vortex charge group is then
\beq
\Z_2=\frac{\Z_4\times\Z_2}{\{(0,0),(1,1),(2,0),(3,1)\}}\ .
\eeq

One may then ask whether the $\Z_2$ charges themselves are conserved.  The W bosons appear to conserve them, but monopole pair creation may not.  A charge 1 monopole is confined by $r_1$ 4-strings and $r_2$ 2-strings, for a total charge of $r_1+r_2\in\Z_2$.  Therefore monopoles are confined by a vortex with nontrivial charge in the remaining $\Z_2$ when $r_1=r_2+1$ mod 2, and so in this case all Douglas-Shenker strings are unstable and monopoles are confined into charge 2 magnetic baryons, which in turn may be bound to W bosons or other monopoles if it is confined by the $\Z_4$ group that was broken by the W bosons.  On the other hand if $r_1=r_2$ mod 2 then the $\Z_2$ string quantum number is preserved.  For example in the vacuum $r_1=0,$\ $r_2=0$ the monopole is unconfined.  On the other hand in the vacuum $r_1=2,$\ $r_2=0$ the monopole is confined by a charge 2 4-string which carries no $\Z_2$ charge, while a charge 2 monopole is unconfined, as is a bound state of a charge 1 monopole and two W bosons.

This spectrum can be read from the usual exact sequence, again with $\H_1(\Sigma_1)=\Z^3$ but now with the induced inclusion map
\beq
i^1_*:\H_1(\Sigma_1)=\Z^3\longrightarrow\H_1(\R^6\times S^1)=\Z:(a_1,a_2,b)\mapsto 4a_1+2a_2+(r_1-r_2)b.
\eeq
which is onto precisely when $r_1-r_2$ is odd.  Therefore the vortex charge group $\H_1(\R^6\times S^1,\Sigma_1)$ is $\Z_2$ when $r_1-r_2$ is even and $0$ when $r_1-r_2$ is odd, while the particle charge group $\H_2(\R^6\times S^1,\Sigma_1)$ is again $\Z^2$.

\subsection{The General Case}
\begin{figure}[ht]
\begin{center}
\leavevmode \epsfxsize 12 cm \epsffile{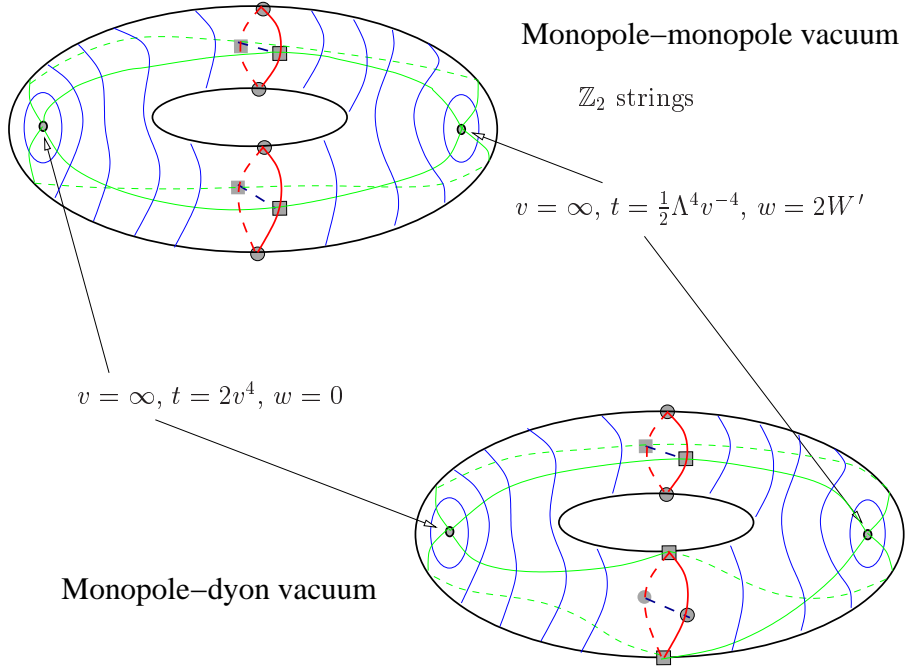}
\end{center}
\caption{ \footnotesize MQCD curves for $U(4) \to U(2) \times
U(2)$. The first represent the monopole-monopole vacuum where the
B-cycle is trivial in $H_1(\R^6\times S^1)$ an so we have
$\mathbb{Z}_2$ Douglas-Shenker string. The second curve represent
the monopole-dyon vacuum where the $B$-cycle wind once in the
M-theory circle and so we have no topological stable string.}
\label{U(4)}
\end{figure}

In general the superpotential may classically break the gauge
symmetry to a product of groups $U(N_i)$, $1\leq i\leq k$.  By now
we have seen that the exact sequence calculation of the
topological charges always leads to a particle charge group \beq
\H_2(\R^6\times S^1,\Sigma_{k-1})=\Z^{2k-2} \eeq where $k-1$ of
the generators are elementary W bosons corresponding to simple
roots of $U(k)$ and the other $k-1$ are the elementary 't
Hooft-Polyakov monopoles.  There can be no torsion elements as
$j^2_*$ is the trivial map and there is no torsion in \beq
\H_1(\Sigma_{k-1})=\Z^{2k-1}. \eeq In addition $\partial^1_*$ is
the zero map and so the group of topological charges carried by
particles is determined by the map \beq
i^1_*:\Z^{2k-1}\longrightarrow\Z:(a_1...a_k,b_1...b_{k-1})\mapsto\sum_i
N_ia_i+(r_{i+1}-r_i)b_i \eeq whose image is \beq
\textup{Image}(i^1_*)=\gcd(N_i,r_{i+1}-r_i)\Z. \eeq This image of
$i^1_*$ is the kernel of $j^1_*$ which determines the group of
particle charges \beq \H_1(\R^6\times
S^1,\Sigma_{k-1})=\frac{\H_1(\R^6\times
S^1)}{\textup{Ker}(j^1_*)}=\frac{\Z}{\gcd(N_i,r_{i+1}-r_i)\Z}=\Z_{\gcd(N_i,r_{i+1}-r_i)}.
\eeq In Figure \ref{U(4)} we see how the B-cycle can affect the
stability of Douglas-Shenker strings.
Notice that the cardinality $t$ of the group $\H_1(\R^6\times
S^1,\Sigma_{k-1})$ is the confinement index, which is the minimal
charge of an unconfined monopole, computed in
\cite{indicediconfinamento}.  The charge $t$ monopole is confined
by $t$ vortices, which carry a charge of $t\in\Z_t$, which is
equivalent to zero modulo $t$ and so the monopole is unconfined.

\section{$\N=1$ $U(N)$ Theory with Flavored Matter} \label{saporisezione}

\subsection{Super QCD: A Global Flavor Symmetry}

In general the inclusion of matter leads to an instability of all Douglas-Shenker strings and often to the existence new, nontorsion, stable vortices.  However we will see that sometimes theories with a gauged flavor symmetry can have stable Douglas-Shenker strings in certain vacua in which all bifundamental quarks have a bare mass.

We begin with the simplest case of fundamental matter that transforms in the fundamental representation of a global $U(N_f)$ flavor symmetry.  The corresponding $\N=2$ theory is described by the Seiberg-Witten curve (\ref{neuno}) which again describes an M5-brane extending along the gauge theory directions $\R^{1,3}$ and a Riemann surface $\Sigma\subset M$.  In the IIA reduction this corresponds to the addition of $N_f$ semi-infinite D4-branes, which we will call flavor branes, that extend from the NS5$\p$ to $x^6=+\infty$.  If two flavor branes are coincident then there is an enhanced flavor symmetry, but arbitrarily small perturbations will separate the branes at sufficiently large $x^6$ and so a classification of stable configurations requires that all of the bare quark masses, that is all of the flavor brane positions on the $v$-plane, are distinct.

Softly breaking the supersymmetry to $\N=1$ with a superpotential, there is now a choice of two supersymmetric configurations for each of the $N$ color D4-branes that extended between the NS5's before the deformation.  Each color brane may either move along $v$ to an extremum of the superpotential as in the previous section, or else it may connect to a flavor brane.  We will ignore configurations in which a color brane touches both a flavor brane and the NS5$\p$.  If a color brane located at some $v$ is connected to a flavor brane on one end then the two NS5's no longer need to be coincident in the $w$-plane, as the D4-branes at $v$ no longer connect them.  The minimum distance between the NS5$\p$ and the locked color-flavor D4-brane pair is the VEV of the meson field corresponding to mesons built from quarks that extend between the now locked color and flavor brane.  These meson VEVs Higgs the color symmetry on the locked branes, and so the remaining classical gauge symmetry group $U(N_1)\times...\times U(N_k)$ has a number of components equal to the number of minima of the superpotential at which unlocked color branes reside.  Thus $k$ again is the number of tubes of the Riemann surface which connect the two NS5's, and so $k-1$ is again the genus of $\Sigma$.

While the genus of the Riemann surface is the same as in the unflavored case, the number of punctures when one deletes the points at infinity \cite{Mikhailov} is now $N_f+2$ instead of just $2$ as in pure super Yang-Mills.  This changes the first homology group of the Riemann surface to
\beq
\H_1(\Sigma)=\Z^{2k+N_f-1} \label{sapori}
\eeq
where the new $N_f$ generators are loops that circle the flavor branes.  The inclusion map $i^1_*$ multiplies these new generators by the number of coincident flavor branes at each bare mass, which is one.  Therefore the inclusion map is onto and again $j_*^1$ is the zero map,
\beq
i^1_*(a_1...a_k,b_1...b_{k-1},c_1...c_{N_f})=\sum_i N_ia_i+(r_{i+1}-r_i)b_i+c_i,\qquad j^1_*=0.
\eeq
However the triviality of $j^1_*$ no longer implies that the group of vortex charges is trivial, because as we will now explain $j^1_*$ is no longer onto as $\partial^1_*$ is no longer necessarily the zero map.

The fact that flavor branes extend to $x^6=+\infty$ with $x^7/x^6\rightarrow 0$ gives a fixed notion of the $x^6$ direction, and so in particular a relative rotation of the two NS5-branes on the $x^6-x^7$ plane may no longer be absorbed into a coordinate redefinition.  Correspondingly in the presence of fundamental matter a Fayet-Iliopoulos (FI) term may no longer be eliminated by a field redefinition.  The FI term, $r$,  corresponds to the $x^7$ coordinate of the NS5$\p$.  All of the color D4's extend in the $x^6$ direction but must remain at constant $x^7$ to preserve supersymmetry, otherwise they would not be parallel to the flavor D4-branes.  Thus when $r\neq 0$ all of the color branes extend from the NS5-brane to $x^6=+\infty$ without touching the NS5$\p$, and so every color is locked.  This does not imply that all flavors are locked, unlocked flavors remain at a constant $x^7=r$ as they extend from the NS5$\p$ to $x^6=\infty$.  As every brane occupies a constant $x^7$ position, if the two NS5's are at different $x^7$ positions then the entire configuration is disconnected.  In general a high energy gauge symmetry consisting of $j$ gauge group components may be engineered using $j+1$ NS5's and if the FI terms are independent then the entire configuration will consist of $j+1$ independent components.

We will now restrict attention to configurations with only two NS5's and so at most two components.  Thus $\H_0(\Sigma)$ will be $\Z^2$ if all colors are locked, but if any color is unlocked then the corresponding D4 connects the two components of the M5 and so $\H_0(\Sigma)=\Z$.  The spacetime is still connected and so $\H_0(\R^6\times S^1)=\Z$ as in pure super Yang-Mills.  This means that the map $i^0_*$ is no longer into when all colors are locked, instead $i^0_*=(1,1)$, and so as suggested above $\partial^1_*\neq 0$.  On the contrary the image of $\partial^1_*$ is the kernel of $i^0_*$, which is the additional $\Z$ component.  This $\Z$ must then appear in its domain, the group of vortex charges
\beq
\H_1(\R^6\times S^1,\Sigma)=\left\{\begin{array}{ll}
\Z\qquad\textup{if all colors are locked}\\
0\qquad\textup{if any color is unlocked.}\\
\end{array}\right.
\eeq

Thus while the Douglas-Shenker vortices may decay via the nucleation of fundamental matter, new nontorsion vortices arise in some cases.  Such vortices have been studied for example in Refs.~~\cite{HT,tong-monopolo,HT2,SY-vortici,SakaiVort}, where it was seen that monopoles in a theory with an FI term are confined by two vortices.  This means that such a vortex cannot decay by monopole-antimonopole nucleation as monopoles are merely kinks in vortex worldvolumes.   Rather than terminating a vortex, these kinks transmute a vortex into a different type of vortex that carries the same conserved charge.

\begin{figure}[ht]
\begin{center}
\leavevmode \epsfxsize 12 cm \epsffile{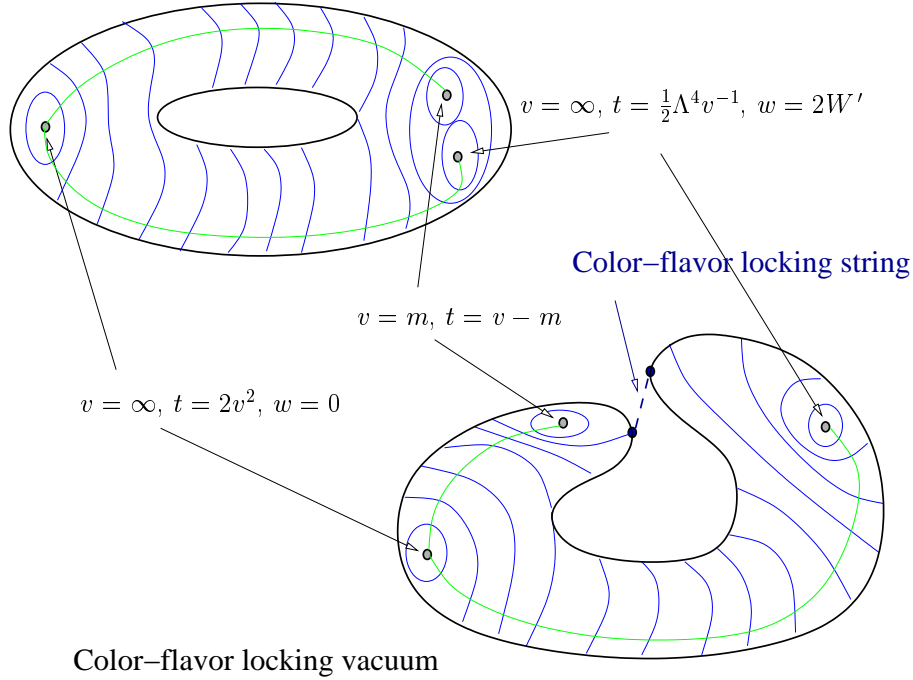}
\end{center}
\caption{\footnotesize MQCD curve for $U(2)$ gauge theory with one
flavor. The second curve represent the color-flavor locking
vacuum.} \label{flavortop}
\end{figure}

The vortices studied in
Refs.~~\cite{Yung:2000uy,VY,MY,vortici,monovortice,stefano}, in
theories with a superpotential rather than an FI term, are not
topologically stable as some of the colors are unlocked.  As a
result there are monopoles confined by a single vortex whose pair
creation causes these vortices to decay (In Figure \ref{flavortop}
we have an example of this kind).
 This does not imply that an FI
term is required for stability. The vortices of those theories
would have been stable had one not included the unlocked color, in
which case the 1-monopoles would have been removed from the
spectrum. Alternately one could have added a new flavor with a
different bare mass and changed the adjoint scalar VEV of the
unlocked color to the bare mass of the new flavor, locking them
together.  The resulting 't Hooft-Polyakov monopoles would have
been confined by two vortices and thus would not have led to
magnetic screening. Note that unlike the vortices of
Ref.~\cite{HT}, such vortices are not be BPS, but the topological
charge is nonetheless conserved. In conclusion, matter that
transforms in the fundamental of a global flavor symmetry screens
Douglas-Shenker string charge, but when the M5-brane is
disconnected a new kind of vortex is stable.

The group of particles is also augmented as a result of the extra $\Z^{N_f}$ in Eq.~(\ref{sapori}).  As the dimension of $\H_1(\R^6\times S^1)$ is still $\Z$, the extra dimensions must come from the group of particle charges
\beq
\H_2(\R^6\times S^1,\Sigma)=\Z^{2k+N_f-2}.
\eeq
The extra $N_f$ independent particles are the quarks with a given color.  A quark of a different color will not carry an independent charge, as its charge is carried by a boundstate of a quark of the given color and a W boson.

\subsection{A Local Flavor Symmetry}

The Douglas-Shenker strings of super QCD are unstable because each quark is defined by a single string, and so a pair of quarks produced on the string leads to a gap which grows until the string has disappeared.  Topologically this reflects the fact that $i^1_*$ is onto because each flavor brane wraps the M-theory circle once and so $i^1_*$ multiplies the quark charges by one.

The situation is quite different if the flavor symmetry is gauged and confining.  In this case there will be two species of Douglas-Shenker strings, those of the color group and those of the flavor group.  A quark will be confined by one of each, and so quark pair-productions, like W boson pair-production in SYM, will only cause vortices to be transmuted into different types of vortices rather than to disappear altogether.  For example a confining $U(N)$ color symmetry and $U(M)$ local flavor symmetry lead to $\Z_N$ charged and $\Z_M$ charged vortices.  A $\Z_N$ vortex becomes a $\Z_M$ vortex when one crosses a quark, and so in particular the charge $M$ $\Z_N$ vortex is unstable.  As in the case of $U(M)\times U(N)$ super Yang-Mills, the group of conserved charges is then
\beq
\H_1(\R^6\times S^1,\Sigma_0)=\Z_{\gcd(M,N)}
\eeq
where $\Sigma_0$ is the lift of three NS5-branes, one pair connected by $N$ D4's and the other by $M$ D4's.  Topologically this is a result of
\beq
i^1_*(a_1,a_2)=Ma_1+Na_2
\eeq
whose cokernel is $\Z_{\gcd(M,N)}$, just as in the case of pure SYM with low energy gauge group $U(M)\times U(N)$ where the W bosons play the role played by the quarks here.

This is not to suggest that gauging the flavor symmetry always stabilizes some of the Douglas-Shenker strings.  For example if any classically unbroken gauge subgroup $U(N_c)$ is in its baryonic root vacua then locally $\Sigma$ factorizes into two funnels which intersect at $2N_c-N_f$ distinct points.  The loop which begins at one intersection, travels along one sheet to the next, and then returns along the other sheet wraps the M-theory circle precisely once.  Thus the image of the corresponding generator of $\H_1(\Sigma)$ will be the element $1\in\H_1(\R^6\times S^1)$.  This means that $i^1_*$ is again onto and no vortices carry any conserved topological charge.  The confined particle must correspond to an M2 that has a boundary that encircles this loop, although it may then encircle a second loop which gives it a net winding number around the M-theory circle of zero.  If the net winding number is zero then the M2 could have disk topology which would allow the particle to be a hypermultiplet \cite{HY,Mikhailov}.  This theory contains flavored magnetic monopoles, which in the Seiberg dual IR free theory become unconfined quarks, and so do not connect to strings.


When the flavor symmetry is gauged, for example in the Klebanov-Strassler theory \cite{KS} in which $x^6$ is periodic, the UV theory may be strongly coupled and in fact potentially continues to cascade.  However the above confined particles exist at each Seiberg duality, as the above cycle may be constructed each time the two M5-brane sheets pass each other.  Thus the vortices confine particles at the energy scale of each step in the cascade, and the vortices will be broken by the lightest particles which come from the last step of the cascade.

\subsection{Gubser-Herzog-Klebanov Axionic Vortices}

Gubser, Herzog and Klebanov have conjectured \cite{GHK} the existence of a new kind of vortex in the Klebanov-Strassler $SU(N+M)\times SU(N)$ theory.  This corresponds to a D1-brane on the conifold which extends in the gauge theory directions.  It does not extend in any internal directions, and so after T-dualizing an internal direction to arrive in IIA it will be a D2-brane extending along the T-dual circle, $x^6$.  Lifting to M-theory one finds an internal spacetime of $\R^5\times T^2$, as in the $\N=4$ example above.  As usual in the IIA reduction there are two NS5-branes, which are at two different coordinates on the $x^6$ circle.

There are now two inequivalent embeddings of the D2-brane extended along $x^6$ and two gauge theory directions.  It may either wrap all of $x^6$, or else it may extend between two NS5's.  In the first case it will be T-dual to a D-string, while in the second it will be T-dual to a half D-string at the conifold singularity which blows up into a D3 when the conifold is resolved.  Both of these branes are non-BPS, as they share three common directions with the D4-branes.  In fact both kinds of brane attract the D4-brane, and once they become coincident will dissolve into the D4.  In the D4 worldvolume theory these are magnetic flux tubes.  The first brane carries a unit of magnetic flux in both gauge groups while the second carries flux in only one.  These flux tubes will smear out of existence unless they are confined for some reason, for example if they really are vortices in an unexpected condensate field in the D4 worldvolume.  In the large $N$ IIB description perhaps the large $N$ limit, which places them far from the horizon's D3-branes that are T-dual to the D4's here, leads to a long lifetime for these strings.

The possibility that these D2's may dissolve into the D4's follows from the topological classification above.  There is no extra $\Z$ component in the group of vortex charges that could stabilize these axionic string charges.  In fact the M5-branes of these models are connected, and so there are no vortices at all carrying topological charges of the type classified in this note.

At distance scales much smaller than the size of the smallest $A$-cycle, corresponding to the $SU(K)$ theory at the bottom of the cascade, one may ignore the D4 and so miss the instability of these vortices.   The decay is caused by the nucleation of degrees of freedom from the $SU(K+M)\times SU(M)$ UV completion and so involves the nucleation of particles of mass equal to the UV cutoff.  Similarly the characteristic lifetime of the vortices will be exponential in the UV cutoff.  This is in contrast with the case of ordinary $SU(K)$ pure super Yang-Mills, which is asymptotically free.  The $SU(K)$ theory at the bottom of the cascade is not asymptotically free.  The fact that both $SU(K)$ theories are in the same universality class does not preclude the existence of stable vortices in one theory and not the other, as the topological stability of a configuration depends on the UV physics.

\section{Conclusions}
In field theories that can be engineered from M5-branes the relative homology groups of the embedding of the M5 yield conserved charges.  In particular we have reproduced the confinement index formula of Ref.~\cite{indicediconfinamento} and we have found the charges corresponding to torsion Douglas-Shenker strings as well as BPS Hanany-Tong strings.  Strings obtained, for example, by softly breaking $\N=2$ SQCD to $\N=1$ with a mass for the adjoint chiral multiplet are found, as expected, to be unstable.  However the conservation of the homology charge led us to conjecture that such vortices may be stabilized by adding a correction to the superpotential that leads the corresponding M5 to be disconnected.  Further we have seen that the M5 always is disconnected, and so the vortices are always stable, in supersymmetric backgrounds with a nonzero FI term and any superpotential.  In both the superpotential and FI cases the disconnected M5 only preserves supersymmetry if all colors are locked to flavors.

Noticably absent from the spectrum of stable vortices is the GHK axionic string.  We have T-dualized the conifold realization of this string to a IIA brane cartoon.  In doing so we have found that there are in fact two species of this string and that both can be continuously deformed into the M5-brane, where no charge that we have seen prevents them from smearing into oblivion.

The topological charge construction considered here, and first proposed in Ref.~\cite{witten-n=1}, applies to a wide variety of theories.  For example the index of confinement can easily be calculated for theories with various kinds of fundamental and bifundamental matter.  Usually the index will be trivial in these cases.  However we have seen that in some theories in which the flavor group is gauged and confining some of the Douglas-Shenker strings are stable and so the confinement index will not be trivial.

The construction itself is much more general than this, allowing treatment of gauge theories in various numbers of dimensions and even of higher (lower) form gauge theories as in the $\N=4$ example above.  The only critical assumption was the validity of the supergravity approximation, and in particular the classical treatment of the geometry.  This is a weaker condition than the supersymmetry of the configuration, for example we have found non-BPS torsion strings and also dyons that cannot satisfy the BPS bound.  However we have not found a criterion that allows one to determine when this approximation applies.

One may attempt to use relative homology to classify the charges of objects in the presence of a domain wall, which is a 3-dimensional cobordism that interpolates between two M5 embeddings $\Sigma$.  Objects with a boundary on the domain wall will be bound to the wall.


\section* {Acknowledgement}

The work of JE is partially supported by IISN - Belgium (convention 4.4505.86), by the ``Interuniversity Attraction Poles Program -- Belgian Science Policy'' and by the European Commission RTN program HPRN-CT-00131, in which he is associated to K. U. Leuven.  Stefano is supported as well.

\end{document}